# Update Query Time Trade-off for Dynamic Suffix Arrays [*] [**]

Amihood Amir and Itai Boneh

Department of Computer Science, Bar-Ilan University, Ramat Gan, Israel,
amir@esc.biu.ac.il, Barbunyaboy2@gmail.com

**Abstract.** The Suffix Array $SA(S)$ of a string $S[1\ldots n]$ is an array containing all the suffixes of $S$ sorted by lexicographic order. The suffix array is one of the most well known indexing data structures, and it functions as a key tool in many string algorithms.
In this paper, we present a data structure for maintaining the Suffix Array of a dynamic string. For every $0 \leq \varepsilon \leq 1$, our data structure reports $SA[i]$ in $\tilde{\mathcal{O}}(n^{\varepsilon})$ time and handles text modification in $\tilde{\mathcal{O}}(n^{1-\varepsilon})$ time. Additionally, our data structure enables the same query time for reporting $iSA[i]$, with $iSA$ being the Inverse Suffix Array of $S[1\ldots n]$.
Our data structure can be used to construct sub-linear dynamic variants of static strings algorithms or data structures that are based on the Suffix Array and the Inverse Suffix Array.

## 1 Introduction

The *suffix tree* [47] and *suffix array* [37] have been, arguably, the most powerful and heavily used tools in Stringology. The *suffix tree* of string $S$ is a compressed trie of all suffixes of $S$, and the *suffix array* of $S$ corresponds to a pre-order traversal of all the leaves of the suffix tree of $S$.

The natural application of the suffix tree is for indexing, but it has been used for many purposes. An incomplete list includes approximate matching [33, 34], parameterized matching [12, 13, 11, 25, 35], efficient compression [50, 51, 2, 48, 22, 42, 39, 14, 9, 1], finding syntactic regularities in strings [10, 31, 15, 32, 26, 49, 28, 21, 46, 30, 36], and much more.

In the 1990's the active field of *dynamic graph algorithms* was started, with the motive of answering questions on graphs that dynamically change over time. For an overview see [20]. Recently, there has been a growing interest in *dynamic pattern matching*. This natural interest grew from the fact that the biggest digital library in the world - the web - is constantly changing, as well as from the fact that other big digital libraries - genomes and astrophysical data, are also subject to change through mutation and time, respectively.

Historically, some dynamic string matching algorithms had been developed. Amir and Farach [6] introduced dynamic dictionary matching, which was later improved by Amir et al. [7]. Idury and Scheffer [27] designed an automaton-based dynamic dictionary algorithm. Gu et al. [24] and Sahinalp and Vishkin [43] developed a dynamic indexing algorithm, where a dynamic text is indexed. Amir et al. [8] showed a pattern matching algorithm where the text is dynamic and the pattern is static. Mehlhorn, Sundar and Uhrig [38] showed how do dynamically maintain a set of sequences while enabling equality queries.

---

[*] This work was partially supported by by ISF grant 1475/18 and BSF grant 2018141.
[**] This work is part of the second author's Ph. D. dissertation.

In the last few years there was a resurgence of interest in dynamic string matching. In 2017 a theory began to develop with its nascent set of tools. Bille et al. [17] investigated dynamic relative compression and dynamic partial sums. Amir et al. [4] considered the longest common factor (LCF) problem. They investigated the case after one error. The fully dynamic LCF problem was tackled by Amir et al. [5], and recently by Charalampopoulos et al. [18]. Gawrychowski at al. [23] used grammars as a tool for maintaining a dynamic collection of strings under various basic operations. Tanimura et al. [40] gave a small space dynamic data structure for longest common extension (LCE) queries.

Throughout all this time, an algorithm for maintaining the suffix tree or suffix array of a dynamically changing text had been sought. The difficulty is that even a single change in the text may cause a linear number of suffixes to change position. Thus, although a dynamic suffix array algorithm would be extremely useful to automatically adapt many static pattern matching algorithms to a dynamic setting, other techniques had to be sought.

Take for example, one of the initial usages for the suffix tree - indexing. Already in 1994, Gu et al. [24] used a data structure construction to allow indexing a dynamic text. Their algorithm can be de-amortized to a $\tilde{\mathcal{O}}(\sqrt{n})$ time for text update and $\tilde{\mathcal{O}}(m\sqrt{n})$ time for an indexing query. This was improved a couple of years later by Sahinalp and Vishkin [43] to just a polylogarithmic slowdown per operation. The powerful idea of Sahinalp and Vishkin was a sophisticated renaming technique. Renaming was also the key to most subsequent efficient dynamic solutions that appeared in the literature.

However, renaming is not a panacea for dynamic algorithms to all the problems that the suffix tree or array solved in the static setting. Perhaps the key property of the suffix array is that the suffixes are **sorted lexicographically**. The powerful renaming and locally persistent parsing techniques developed thus far do not maintain lexicographic ordering. It is, thus, no surprise that problems like maintaining the Burrows-Wheeler transform, or finding the Lyndon word of a substring, do not hitherto have an efficient dynamic version.

The only papers we found in the literature that attempt to compute the suffix array and Burrows-Wheeler transform on a dynamic text are Salson et al. [44, 45]. These algorithms are useful in practice, but their asymptotic worst-case complexity is still **linear** per update. To our knowledge, our paper provides the first algorithm that maintains the lexicographic ordering of suffixes in asymptotic worst-case **sublinear** time.

The contributions of this paper are:

1. We provide the first algorithm for maintaining the suffix array of a dynamic string in sublinear time. For every $0 \leq \varepsilon \leq 1$, our algorithm reports the $i$th entry in the suffix array, $SA[i]$ in time $\tilde{\mathcal{O}}(n^\varepsilon)$ and handles text modification in time $\tilde{\mathcal{O}}(n^{1-\varepsilon})$. Additionally, our algorithm enables the same query time for reporting $iSA[i]$ with $iSA$ being the Inverse Suffix Array of $S[1\ldots n]$.
2. We define a simple and efficient new data structure, which we call the *kTree*. This data structure is what powers the algorithm.
3. Our algorithm provides immediate sublinear algorithms to various important problems for which there is no known dynamic algorithm. Examples are computing the Burrows-Wheeler transform, and finding the Lyndon root of a given text substring.

This paper is organized as follows: Section 2 gives basic definitions and terminology. In Section 3, the k-Words Tree data structure is defined. Section 4 provides the algorithm for the inverse suffix array of a dynamic text, and Section 5 describes and analyses the algorithm for dynamic suffix array maintenance. We conclude in Section 6 with algorithms for problems that had no efficient dynamic algorithms till now, and where such algorithms are immediately derived from our dynamic suffix array algorithm.

## 2 Preliminaries

We begin with basic definitions and notation generally following [19].

Let $S = S[1]S[2]\ldots S[n]$ be a *string* of length $|S| = n$ over a finite ordered alphabet $\Sigma$ of size $|\Sigma| = \sigma = O(1)$. By $\lambda$ we denote an empty string. For two positions $i$ and $j$ on $S$, we denote by $S[i\ldots j] = S[i]\ldots S[j]$ the *factor* (sometimes called *substring*, and sometimes *word*) of $S$ that starts at position $i$ and ends at position $j$ (it equals $\lambda$ if $j < i$). We recall that a *prefix* of $S$ is a factor that starts at position 1 ($S[1\ldots j]$) and a *suffix* is a factor that ends at position $n$ ($S[i\ldots n]$). We denote the *reverse* string of $S$ by $S^R$, i.e. $S^R = S[n]S[n-1]\ldots S[1]$.

Let $Y$ be a string of length $m$ with $0 < m \leq n$. We say that there exists an *occurrence* of $Y$ in $S$, or, more simply, that $Y$ *occurs in* $S$, when $Y$ is a factor of $S$. Every occurrence of $Y$ can be characterised by a starting position in $S$. Thus we say that $Y$ occurs at the *starting position* $i$ in $S$ when $Y = S[i\ldots i+m-1]$. We say that a string $S$ of length $n$ has a *period* $p$, for some $1 \leq p \leq \frac{n}{2}$ if for every $i$ such that $1 \leq i \leq n - p$, it is the case that $S[i] = S[i+p]$. *The period* of $S$ is the smallest $p$ for which that condition holds.
The *Concatenation* of two strings $S[1\ldots n]$ and $T[1\ldots m]$ denoted as $S \cdot T$ or simply as $ST$ is the string generated by appending $T$ to the end of $S$. Namely, $S \cdot T = S[1]S[2]\ldots S[n]T[1]T[2]\ldots T[m]$.
We say that a substring of $S$, denoted as $A = S[a\ldots b]$ is a *run* with period $p$ if its period is $p$, but $S[a-1] \neq S[a-1+p]$ and $S[b+1] \neq S[b+1-p]$. This means that no substring containing $A$ has a period $p$.
Let $A = S[i\ldots j]$ be a substring of a text $S$ and let $1 \leq x \leq n$ be an index in $S$. We say that $x$ *is touching* $A$ if $x = i - 1$ or $x = j + 1$. We say that $x$ *is contained within* $A$ if $i \leq x \leq j$. Let $B = S[i_B\ldots j_B]$. We say that $B$ is *contained within* $A$ if every index in $B$ is contained within $A$, equivalently, $B$ is *contained within* $A$ if $i \leq i_B \leq j_b \leq j$. $B$ is said to be *strongly contained* within $A$ if $i < i_B \leq j_B < j$.

Given two strings $S$ and $T$, the string $Y$ that is a prefix of both is the *longest common prefix (LCP)* of $S$ and $T$ if there is no longer prefix of $T$ that is also a prefix of $S$.

**Longest Common Extension Queries** are a useful tool in string algorithms. Given a text $T$, the *longest common prefix (LCP)* of two indices $i$ and $j$, denoted as $LCP(i, j)$, is the longest substring that is the prefix of both of the suffixes $T[i \ldots n]$ and $T[j \ldots n]$. The *longest common suffix (LCS)* of $i$ and $j$, denoted as $lcs(i, j)$, is the longest common suffix of $T$'s prefixes ending in $i$ and $j$. LCP and LCS queries are called longest common extension queries. Longest common extension queries in a static string can be answered in constant time following a linear time preprocessing (for finite fixed alphabets. There are logarithmic or $\log \log$ factors for various infinite alphabets) [16]. In a dynamic string, the following holds:

**Lemma 1.** *Given a dynamic text $T$, there is a data structure for answering dynamic longest common extension queries in polylogarithmic time. Maintaining this data structure takes polylogarithmic time per substitution.*

The above result has been continually improved by a list of papers [38, 41, 23], culminating in the most efficient deterministic algorithm for longest common extension in a dynamic text, that of Nishimoto et al. [40].

The **suffix array** of a string $S$, denoted as $SA(S)$, is an integer array of length $n + 1$ storing the starting positions of all (lexicographically) sorted non-empty suffixes of $S$, i.e. for all $1 < r \leq n + 1$ we have $S[SA(S)[r-1] \ldots n] < S[SA(S)[r] \ldots n]$. Note that we explicitly add the empty suffix to the array. The suffix array of $S$ corresponds to a pre-order traversal of all the leaves of the suffix tree of $S$. The inverse $iSA(S)$ of the array $SA(S)$ is defined by $iSA(S)[SA(S)[r]] = r$, for all $1 \leq r \leq n + 1$. Let $S$ and $R$ be strings, we denote the lexicographic order between them by $<_L$ or $\leq_L$, i.e. $S <_L R$ means *S is lexicographically smaller than R*.

Let $S_1, \ldots, S_k$ be strings over alphabet $\Sigma$ and let $\$ \notin \Sigma$. We assume that every string $S_i$, $i = 1, \ldots, k$, ends with a $\$$ symbol. An *uncompacted trie of strings* $S_1, \ldots, S_k$ is an edge-labeled tree with $n$ leaves. Every path from the root to a leaf corresponds to a string $S_i$, with the edges labeled by the symbols of $S_i$. Strings with a common prefix start at the root and follow the same path of the prefix, and the paths split where the strings differ. A *compacted trie* is the homeomorphic image of the uncompacted trie, i.e., every chain of edges connected by degree-2 nodes is contracted to a single edge whose label is the concatenation of the symbols on the edges of the chain.

Let $S = S[1], \ldots, S[n]$ be a string over alphabet $\Sigma$. Let $\{S_1, \ldots, S_n\}$ be the set of suffixes of $S$, where $S_i = S[i], S[i+1], \ldots, S[n]$, $i = 1, \ldots, n$. A *suffix tree* of $S$ is the compacted trie of the suffixes $S_1, \ldots, S_n$. We associate every node $V$ is the suffix tree with the string $L(V)$, the concatenation of the strings on the edges from the root to $V$.



## 3 The k-Words Tree

We start by defining a data structure that is fundamental to our algorithm.

**Definition 1.** *Let $S[1\ldots n]$ be a string and let $1 \leq k \leq n$ be an integer. Let $D_k^S$ be the set of all different k-length substrings of S. The k-Words Tree $kT(S)$ of S is a balanced search tree, where every node V represents a word $W(V) \in D_k^S$. $W(V)$ will be referred to as the word of V. $T(V)$ Contains all the indices i such that $S[i\ldots i+k-1] = W(V)$. These are all the instances of the word of V. The indices in $T(V)$ are sorted in increasing order of their numeric value. The nodes of $kT(S)$ are sorted by lexicographic order of the $W(V)$s. Additionally, every node V maintains auxiliary information about the number of items in the sub-tree rooted in V. By* items *we refer to the overall amount of elements in $T(V)$ over every V in the rooted subtree.*

For an example, see Fig. 1.

S = ABBABABAABBABAABBBAA

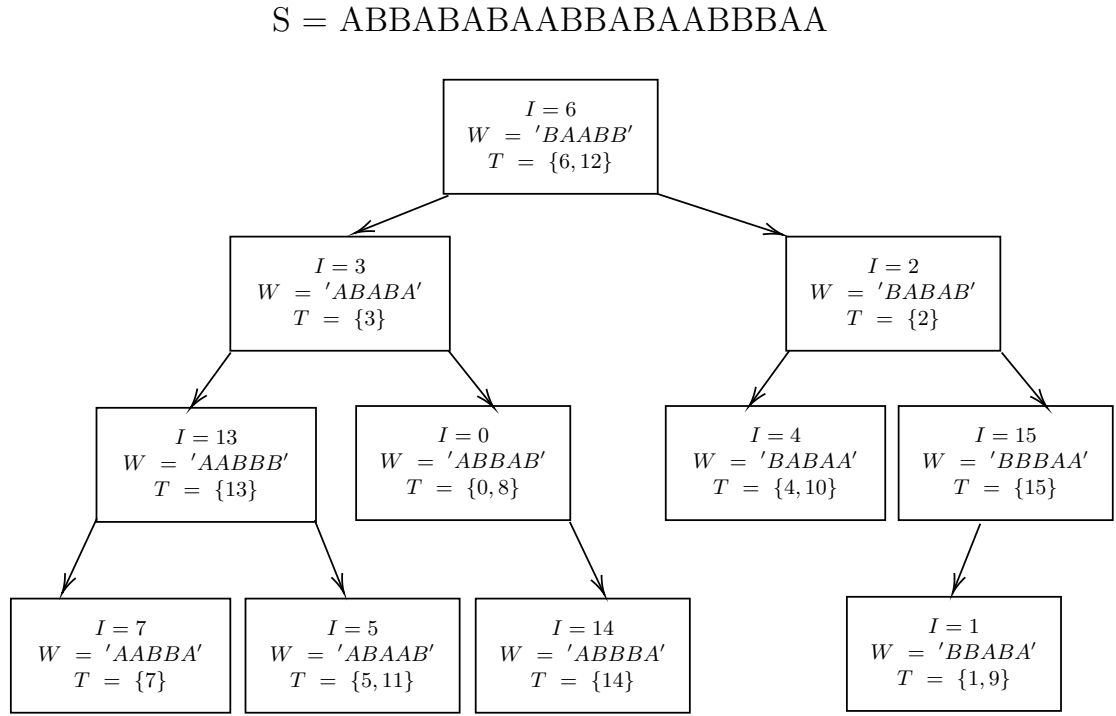

**Fig. 1.** An Example for a k-Words tree with $k = 5$. $W(V)$ **is not** explicitly stored in V and is only specified for clarification .The auxiliary information is omitted



In order to define a $k$-length substring starting at every $1 \leq i \leq n$, we append $\$^k$ at the end of $S$, where the symbol $\$$ does not belong to $\Sigma$ and is lexicographically greater than every $\sigma \in \Sigma$.

Given a string $S[1\ldots n]$, $kT(S)$ can be constructed in $\tilde{\mathcal{O}}(n)$ time as follows: Preprocess $S$ for constant time LCP queries. This enables lexicographic comparisons between suffixes of $S$ in constant time. Given two suffixes $i$ and $j$ we can check $LCP(i,j) = l$ and than compare $S[i+l]$ to $S[j+l]$ to decide the lexicographical order of the suffixes.

Initialize an empty balanced search tree $kT(S)$ and for every $1 \leq i \leq n$: search $S[i\ldots i+k]$ in $kT(S)$ using an LCP query in every node to compute the lexicographic comparison. If a node $V$ such that $W(V) = S[i\ldots i+k]$ is met, simply add $i$ to $T(V)$. Otherwise, add a new node to $kT(S)$ with $W(V) = S[i\ldots i+k]$ and $T(V)$ containing $i$.

Equal $k$-length words can be identified by checking if $LCP(i,j) \geq k$. The auxiliary information about the number of nodes in the rooted subtrees is maintained upon changes.

Since LCP queries can be answered in $\tilde{\mathcal{O}}(1)$ time in dynamic settings (Lemma 1), we obtain the following:

**Lemma 2.** $kT(S)$ can be maintained in $\tilde{\mathcal{O}}(k)$ time per symbol substitution update.

**Proof:** Denote the index where the substitution occurred as $x$. The only words that change are the words starting in $i$, $x - k < i \leq x$. Before applying $S[x] = \sigma$, remove all the words that start in $i$, $x - k < i \leq x$ from $kT(()S)$. That is done in $\tilde{\mathcal{O}}(1)$ per word by searching for $S[i\ldots i+k]$ in the balanced search tree $kT(S)$ and removing $i$ from the respective $T(V)$, where the indices appear in increasing order. If $T(V)$ is empty - remove $V$ from $kT(S)$. Either way update the items-count in every node up the route to the root.

After removing all the modified words, we apply $S[x] = \sigma$ and add all the words starting in $x - k < i \leq x$ as in the initialization of $kT(S)$.

Since both inserting and removing a word takes $\tilde{\mathcal{O}}(1)$ time, and we use exactly $k$ removals and $k$ insertions, modifying $kT(S)$ takes $\tilde{\mathcal{O}}(k)$ time. □

The k-Words tree will serve as the basis for our dynamic suffix array algorithm. The main effort in this paper is showing that with appropriate auxiliary data, information on the lexicographic order of suffixes can be derived from the k-Words tree.

**Remark:** In order to generalize Lemma 2 to support deletions and insertions, An additional obstacle needs to be handled. Even though the number of modified words remains $O(k)$, the *starting indices* of the words following the updated index is modified. Namely, after an insertion update in index $i$, every word with a starting index $j > i$ needs to have its index modified to $j + 1$. This can be handled in various ways using basic data structure techniques. We leave the details for the full version of this paper.

## 4 Dynamic Inverse Suffix Array

### 4.1 The Idea

Our goal is to compute $iSA[i]$. In other words, given an index $i$ - we wish to find the number $x$ s.t. $SA[x] = i$. This is equivalent to reporting the number of suffixes that are lexicographically smaller than $S[i \ldots n]$.

We call two suffixes $S[i \ldots n], S[j \ldots n]$ *close* suffixes if $LCP(i, j) \geq k$. Otherwise, $i$ and $j$ are *far* suffixes. Note that the starting indices of far suffixes are contained within the $T(V)$ of different nodes of the k-Words tree, while the starting indices of close suffixes are contained within the $T(V)$ of the same node in the k-Words tree.

**Lemma 3.** *The number of far suffixes that are lexicographically smaller than $S[i \ldots n]$ can be obtained from the k-Words tree in time $\tilde{\mathcal{O}}(1)$.*

**Proof:** Given an input index $i$, we transverse from the root of the k-Words tree towards the node containing $S[i \ldots i+k]$. This route requires an LCP query in every node, since we need to determine whether $S[i \ldots n]$ is lexicographically smaller or larger than the word represented by the current node. Every time we go to the right, all the items in nodes to the left correspond to far suffixes that are smaller than $S[i \ldots n]$. We accumulate the sum $C$ of the items contained in the nodes to the left of our route using the auxiliary data. Once the node $V$ containing $i$ is met, we add the number of items within the subtree rooted in the left child of $V$ to $C$. By the end of this route, $C$ is the amount of items to the left of $V$ in $kT(S)$, which is exactly the number of suffixes that are far from $S[i \ldots n]$ and lexicographically smaller than $S[i \ldots n]$. □

We are left with the task of counting the *close* suffixes that are lexicographically smaller than $i$. The starting indices of the *close* suffixes are exactly the indices stored in $T(V)$. It is possible that there are a lot of *close* suffixes, but since these indices correspond to instances of a word of length $k$, periodicity can be exploited in order to implicitly compare $S[i \ldots n]$ to **all** of the *close* words in $\tilde{\mathcal{O}}(\frac{n}{k})$ time.

### 4.2 Counting Smaller Close Suffixes

The close suffixes are listed in $T(V)$ with $V$ being the node containing $S[i \ldots i+k]$. We need to find how many of them are lexicographically smaller than $S[i \ldots n]$. We can not explicitly compare $S[i \ldots n]$ to all of them since there may be too many. The following observation follows from Fact 5 in [3] and it is the key for handling all the (possibly $O(n)$) suffixes in $\tilde{\mathcal{O}}(\frac{n}{k})$ time.

**Observation 1.** *Let $i_1, i_2, \ldots, i_s$ be the set of indices in $T(V)$ in increasing order. And let $i_s, i_{s+1}$ be two adjacent indices. If $i_{s+1} - i_s < \frac{k}{2}$, then $i_s$ is contained in a run $R = S[a \ldots b]$ of length at least $k$ with a period $p = i_{s+1} - i_s$. The next consecutive indices are $i_p = i + p \cdot t$ for $i_p \leq b - k$. The next index after this set is at least $\frac{k}{2}$ larger than its predecessor.*

Observation 1 can be used to identify *periodic clusters* of indices. If we transverse on the values in $T(V)$ in ascending order, and encounter two adjacent indices s.t. $i_{s+1} - i_s < \frac{k}{2}$, we can find the run $R = S[a \ldots b]$ by querying $LCS[i_s, i_{s+1}]$ - this will yield the extension of the run to the left from $i_s$. We represent all the indices that are contained within the run as an arithmetic progression and proceed to the successor of $i_l$ in $T(V)$, $i_l$ being the greatest item in the arithmetic progression. According to the Observation 1, that will result in a value that is at least $\frac{k}{2}$ larger than $i_l$. Notice that every step in this iteration results in the next element (either an index or cluster) starting in an index greater than the previous one by at least $\frac{k}{2}$. Since the largest possible index is $n - k$, this iteration terminates within at most $O(\frac{n}{k})$ steps, and yields a representation of size $O(\frac{n}{k})$ of all the indices in $T(V)$. We denote such a periodic cluster by $C[a, b, p]$, with $a$ being the index of leftmost occurrence of $W(V)$ within the periodic run, $b$ being the ending index of the run, and $p$ being the period of the run.

Note that the indices in $T(V)$ are stored explicitly, and not as periodic clusters. Therefore, no further treatment is necessary to maintain them in cluster form upon an update. The cluster representation is generated from the indices during the query execution for the relevant node.

By now, we showed how to obtain a representation of all the occurrences of $W(V)$, corresponding to the starting indices of all the close suffixes, in $\tilde{\mathcal{O}}(\frac{n}{k})$ time. The representation consists of $\tilde{\mathcal{O}}(\frac{n}{k})$ elements, some are single indices and some are clusters. So the remaining challenge is to compare all the suffixes corresponding to the indices within a cluster to $S[i \ldots n]$ in polylogarithmic time.

**Lemma 4.** *Given an index $i$ and a periodic cluster of indices $C = [a, b, p]$, the amount of indices within $C$ that correspond to suffixes that are lexicographically smaller than $S[i \ldots n]$ can be calculated in time $\tilde{\mathcal{O}}(1)$.*

**Proof:** Let $r_i$ and $r_a = b - a$ be the extensions to the right of the run with period $p$ from indices $i$ and $a$ respectively. $r_i$ can be calculated by an LCP query $LCP(i, i + p)$. Let $i_t = a + t \cdot p$ be an index within $C$. The result of the lexicographic comparison between the indices of $C$ and $i$ can be partitioned into three classes (for better intuition, see Figure 2:

1. $r_i < r_a - t \cdot p$ : In this case, the mismatch between the two suffixes will occur between index $i + r_i$ in suffix $S[i \ldots n]$ and index $i_t + r_i$ in suffix $S[i_t \ldots n]$. $i + r_i$ is independent of $t$. $i_t + r_i$ is within the run of period $p$ because $i_t + r_i < i_t + r_a - t \cdot p = a + r_a$. Additionally $i_t + r_i$ is always the same mod $p$. So for every $i_p$ in this case, $S[i_t + r_i]$ is the same symbol. Therefore, the result of comparing between $i$ and every $i_t$ in this case is determined by the result of comparing between $S[i + r_i]$ and $S[i_t + r_i]$. For some arbitrary $i_t$ that satisfies the condition for this case.
2. $r_i > r_a - t \cdot p$ : symmetrically to the first case, the mismatch is between index $i + r_a - t \cdot p$ in the suffix $S[i \ldots n]$ and index $a + r_a$ in the suffix $S[i_p \ldots n]$. Symmetric reasoning leads to $S[i + r_a - t \cdot p]$ being the same symbol for every $t$ within this case. Again, all the comparison results are determined by a single symbol comparison.

3. $r_i = r_a - t \cdot p$ : That case can occur for at most one index in $C$. This index can be calculated in constant time. Since there is only a single element in this case - we can treat it explicitly with a single $LCP$ query.

We showed how to handle each case using a constant amount of $LCP$ queries, so the overall time for counting the suffixes smaller than $S[i \ldots n]$ within the cluster is $\tilde{\mathcal{O}}(1)$ □

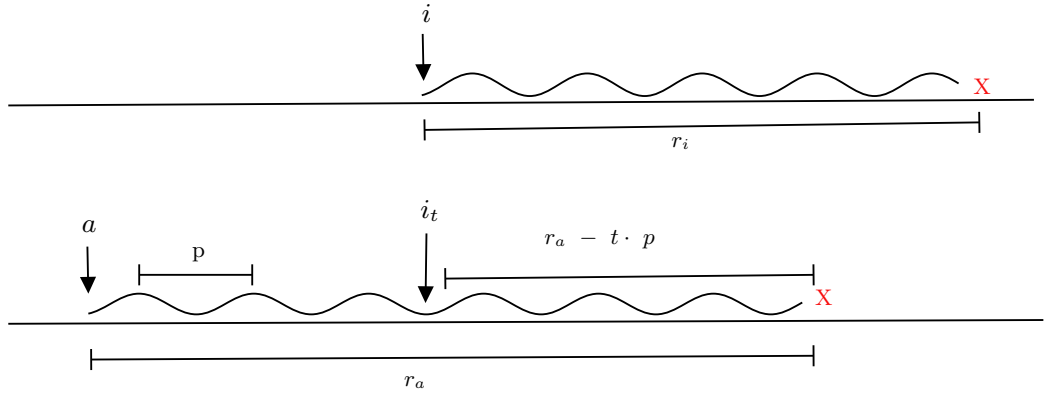

**Fig. 2.** The settings of Lemma 4. Specifically case 2. Red 'X' represents the ending of a run.

With Lemma 4, our algorithm is completed. We start by applying Lemma 3 to count the far suffixes that are lexicographically smaller than $S[i \ldots n]$ and, as a side effect, find the node $V$ in the k-Words tree that contains $i$. We extract the compact representation of all the close suffixes from $T(V)$ in $\tilde{\mathcal{O}}(\frac{n}{k})$ time. For every element (either a single occurrence or a cluster), compare the corresponding suffix (or suffixes) to $S[i \ldots n]$ by using either an LCP query (for a single occurrence) or Lemma 4 (for a cluster). When the iteration is completed, we have the number of lexicographically smaller *close* suffixes. Since comparing $S[i \ldots n]$ to either a single occurrence or a cluster takes $\tilde{\mathcal{O}}(1)$ time, this iteration takes $\tilde{\mathcal{O}}(\frac{n}{k})$ time overall. We output the sum of close smaller suffixes and the far smaller suffixes.

## 5  Dynamic Suffix Array

Finding $SA[i]$ is equivalent to finding the suffix that is lexicographically larger than **exactly** $i$ other suffixes. The idea is similar to the idea for the inverse suffix array. First, use $kT(S)$ to find the node $V$ that must contain $SA[i]$. The second step is identifying which one of the indices in $T(V)$ is $SA[i]$.

### 5.1 Finding The Containing Node

Node $V$ in $kT(S)$ that contains the suffix that is lexicographically greater than exactly $i$ other suffixes, is found by using the following recursive procedure.
**LargerThan(Root, i)**:

1. Denote the number of items in the subtree rooted in the left child of $Root$ as $|L|$. If $i < L$: return **LargerThan(LeftChild , i)**
2. If $|L| \geq i$ and $i < |L| + |T(Root)|$ then return $(Root, i - |L|)$.
3. If $|L| + |T(Root)| \geq i$ return **LargerThan(RightChild , $i - (|L| + |T(Root)|)$)**

The procedure takes logarithmic time since every recursive call is for a lower child in $kT(()S)$, which is a balanced search tree.

Notice that the algorithm returns an index as well. This is the lexicographic rank of $SA[i]$ among the suffixes corresponding to the indices of $T(V)$.

### 5.2 Finding SA[i] in $T(V)$

With the identification of $V$, the problem is reduced to finding the $i$th lexicographically smallest element in $T(V)$. A standard approach would be sorting $T(V)$ by the lexicographic order of its elements, but that can not be done without unpacking the compact representation of $T(V)$'s elements. A single periodic cluster of indices does not necessarily form a consecutive block in the lexicographic sorting of $T(V)$, so it is possible that the sorted $T(V)$ is not representable in $O(\frac{n}{k})$.

Our approach is using the routine from the $iSA$ algorithm for finding the position of an element $x$ in the suffix array. If we run this routine on some item $x$, and find that $iSA[x] < i$, we can recursively proceed on all the elements that are lexicographically larger than $x$. If we manage to find an element $x$ that is sufficiently close to being the lexicographic median of the remaining elements, then this will be similar to a binary search.

**Definition 2.** *Given $A$ a set of suffixes, and let $0 < \alpha < \frac{1}{2}$ be a real number. $x \in S$ is an $\alpha$-good pivot for $A$ if at least $\alpha \cdot |A|$ suffixes in $A$ are lexicographically smaller than (or equal to) $x$ and at least $\alpha \cdot |A|$ suffixes in $A$ are lexicographically larger than (or equal to) $x$.*

The following observation is the key for efficiently finding a $\frac{1}{4}$-good pivot:

**Observation 2.** *Every cluster of instances $C = [a, b, p] = a, a+p, a+2p\ldots$ is either a lexicographically decreasing or a lexicographically increasing sequence of suffixes.*

**Proof:** Consider two consecutive suffixes $i, i+p$. The mismatch between these two suffixes is in the first index where the run halts for one of them. That is index $b+1$ for the suffix $S[i+p\ldots n]$ and index $b+1-p$ for the suffix $S[i\ldots n]$. These two indices are independent from $i$, so the lexicographic order of every two consecutive suffixes in $C$ is the same. □

Armed with Observation 2, we find an $\frac{1}{4}$-good pivot as follows: Compute the periodic cluster representation of size $O(\frac{n}{k})$ for $T(V)$. For every cluster $C = [a, b, p]$ calculate $m_C$ the middle term in the arithmetic progression corresponding to $C$. According to Observation 2, that is the lexicographic median of the cluster. $m_C$ can be obtained via simple arithmetic operations on $[a, b, p]$.

Now we have a set of all the cluster medians and all the non-cluster elements in $T(v)$. We sort them lexicographically using a classic comparisons sorting algorithm to obtain a lexicographically sorted array $A = i_1, i_2 \ldots i_t$. Initialize a counter $c = 0$ and start iterating $A$ from left to right. For every index: if it is a non-cluster median, then increase $c$ by 1; if it is a cluster median, increase $c$ by $\lceil |C|/2 \rceil$, where $|C|$ is the number of elements in the cluster. Halt in the first element $i_p$ where $c > \frac{S}{4}$, where $S$ is the overall number of elements.

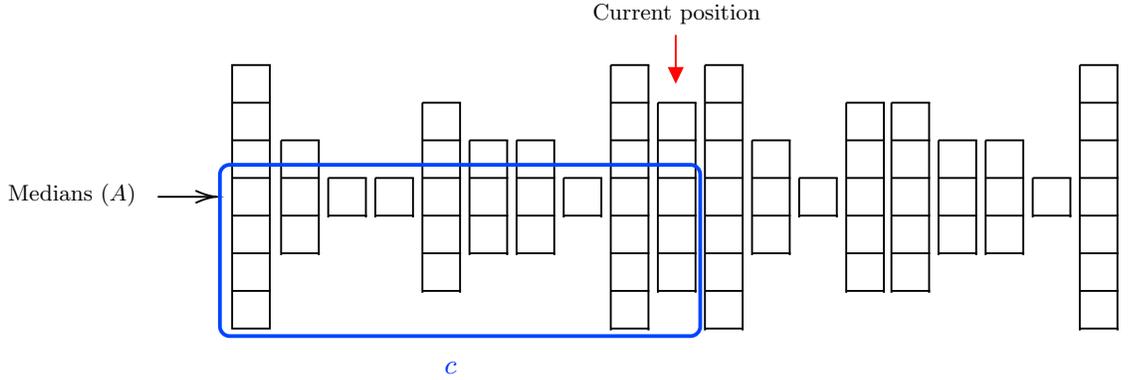

**Fig. 3.** An illustration of the process for finding $i_p$. Every square represents a suffix, and every vertical grid represents a periodic cluster of suffixes. The blue rounded rectangle contains $c$ squares.

*Claim.* The suffix starting in $i_p$ is a $\frac{1}{4}$-good pivot.

**Proof:** Every element in the $A$ is the median of its respective cluster, so it is greater (or equal to) half of the suffixes in its cluster. Since $A$ is a lexicographically increasing array, the suffix $i_{k+1}$ that is to the right of suffix $i_k$ is greater than at least the same number of suffixes as $i_k$, plus half the size of its own cluster, which is added to $c$ when $i_{k+1}$ is visited. It follows that at every point in the iteration, the currently iterated suffix is greater than or equal to at least $c$ other suffixes. Therefore, $i_p$ is greater or equal to at least $\frac{1}{4}|A|$ other suffixes.

Recall that $S$ is the amount of suffixes represented by the clusters and individual suffixes of $A$. Let $C_k$ be the size of the $k$'th cluster in $A$ (if $i_k$ is a single suffix, $C_k = 1$. It's easy to see that $\Sigma_{k=1}^{t} \lceil \frac{C_k}{2} \rceil \geq \frac{|E|}{2}$ and that when $i_a$ is contested in the iteration, $c = \Sigma_{k=1}^{a} \lceil \frac{C_k}{2} \rceil$. Since $i_p$ is the **first** element to have $c > \frac{|C|}{4}$, $c$ was smaller than $\frac{S}{2}$ when $i_{p-1}$ was contested. Namely: $\Sigma_{k=1}^{p-1} \lceil \frac{C_k}{2} \rceil < \frac{S}{4}$. Since the sum of **all** $\lceil \frac{C_k}{2} \rceil$ is at least $\frac{S}{2}$, it follows that $\Sigma_{k=p}^{t} \lceil \frac{C_k}{2} \rceil > \frac{|S|}{4}$. Every median (or single suffix) to the right of $i_p$, including $i_p$, is corresponding to $\lceil \frac{C_k}{2} \rceil$ suffixes that are lexicographically greater (or equal to) $i_p$. These are the suffixes above the respective median including itself. So there are at least $\Sigma_{k=p}^{t} \lceil \frac{C_k}{2} \rceil > \frac{|S|}{4}$ suffixes in that are lexicographically greater than $i_p$. □

We find $SA[i]$ in a binary search fashion. We start by finding $i_p$ and $x = iSA[i_p]$. if $x = i$, then, we are done (output $i_p$). Otherwise, if $x > i$ we reduce the range of our search to the suffixes that are lexicographically larger than $i_p$ and vice versa if $x < i$. Recursively proceed with the remaining suffixes. The suffixes for the next search iteration can be found and efficiently represented using the same method as for finding $iSA[i_p]$. We are guaranteed to eliminate at least $\frac{1}{4}$ of the remaining elements in every iteration of the search, so this procedure will terminate within $O(\log n)$ iterations.

**Time:** Finding the pivot takes $\tilde{\mathcal{O}}(\frac{n}{k})$ for extracting the medians from the clusters and $\tilde{\mathcal{O}}(\frac{n}{k})$ for sorting the array of medians. The recursive procedure for finding $SA[i]$ takes $T(s) = T(\frac{3s}{4}) + \tilde{\mathcal{O}}(\frac{n}{k})$ with $s \leq n$ being the amount of implicitly represented suffixes in the set. This recursive formula is dominated by $\tilde{\mathcal{O}}(\frac{n}{k})$.

## 6 Applications

Our data structure enables a flexible trade-off between the update time and the query time. Setting $k = n^\epsilon$ yields a data structure with $\tilde{\mathcal{O}}(n^\epsilon)$ update time and $\tilde{\mathcal{O}}(n^{1-\epsilon})$ query time for either a suffix array or an inverse suffix array query.

In the following subsections we give examples of problems that can be solved in a dynamic setting by our methods. Some of these are motivated by *internal* string problems as explored by Kociumaka [29].

### 6.1 Substring Lyndon Root

The Lyndon root of a string $S[1 \ldots n]$ is the lexicographically smallest rotation $S[i \ldots n] \cdot S[1 \ldots i-1]$ of $S$. Equivalently: The Lyndon root of $S$ is the lexicographically smallest suffix of the string $S \cdot S$ starting in $[1 \ldots n]$.

With a little bit of extra work we can use the dynamic suffix array to find the Lyndon root of a given substring of $S$. Namely: Given $i < j$ two indices in $S$, output the Lyndon root of $S[i \ldots j]$.

**Observation 3.** *Let $S[1 \ldots n]$ be a dynamic string. The suffixes starting in an interval $I = [a \ldots b]$ can be maintained in $\tilde{\mathcal{O}}(|I|^\epsilon)$ update time to support lexicographic select queries among these suffixes in $O(|I|^{1-\epsilon})$ time. With $|I| = b - a + 1$. Namely, the input query is an index $1 \leq i \leq |I|$ and the output is the i'th lexicographic query of $S$ among the queries starting within $I$.*

**Proof:** By maintaining $kT(S[a\ldots b+k])$, and applying the dynamic suffix array query of section 5. Setting $k = |I|^\epsilon$ yields the desired complexity. □

Note that if we apply Observation 3 to dynamically maintain the lexicographic order of suffixes starting in $I[a\ldots b]$, only updates in indices $i \in [a\ldots b+|I|^\epsilon]$ require applying an update to our data structure.

For the purpose of reporting the Lyndon root of a substring, we dynamically maintain the lexicographical order of the suffixes starting in $S[a\cdot 2^t \ldots a+1)\cdot 2^t - 1]$ for every $0 \le t \le \log(n)$ and $0 \le a < \frac{n}{2^t}$ by applying Observation 3.

Given an update in $S$, only $O(\log(n))$ substring suffix arrays need to be modified, a constant number for every value of $t$. The complexity for updating the lexicographic suffix sort for all the intervals that require an update is $\Sigma_{t=0}^{\log(n)}\tilde{\mathcal{O}}((2^t)^\epsilon)$. Being the sum of a geometric progression, this expression is dominated by the last element $\tilde{\mathcal{O}}(n^\epsilon)$.

Given a substring $S' = S[i\ldots j]$ we can partition this substring to $\log(|S'|)$ intervals of the form $S[a \cdot 2^t \ldots (a+1)\cdot 2^t - 1]$. Next, we adjust the data structure of every interval in the partition to compare between suffixes as if another copy of $S'$ is concatenated to the end of $S'$. In the worst case, this adjustment may require updating all the words in the k-tree that touch the right end of the subword, which takes $\tilde{\mathcal{O}}(k) = \tilde{\mathcal{O}}(|I|^\epsilon)$ for the data structure of interval $I$. $LCE$ queries on $S' \cdot S'$ are required both for applying the adjustment and for executing the query. These can be answered by employing the fact that the LCE data structure of [40] supports cutting and inserting substrings of $S$ in polylogarithmic time.

After adjusting the $k$-trees of the intervals and the LCE data structure, we query every interval for the minimal suffix starting within it. This yields $log(S')$ candidates for being the Lyndon root of $S'$. we return the minimal suffix among them. After returning the Lyndon root, we undo the modifications for the $k$-trees of the intervals and for the LCE data structure.

**Theorem 1.** *For every $0 \le \epsilon \le \frac{1}{2}$, a dynamic string $S[1\ldots n]$ undergoing symbol substitution updates can be maintained so reporting the Lyndon root of $S' = S[i\ldots j]$ can be done in time $\tilde{\mathcal{O}}(|S'|^{1-\epsilon})$ and updates are handled in time $\tilde{\mathcal{O}}(n^\epsilon)$.*

**Proof:** The bottleneck of executing the query is adjusting, querying and undoing the adjustments for every interval in the partition of $S'$. These intervals are of length at most $|S'|$. Updating the $k$-tree for every such interval takes at most $\tilde{\mathcal{O}}(|S'|^\epsilon)$. Querying the intervals will take $\tilde{\mathcal{O}}(|S'|^{1-\epsilon})$ There are $O(\log(|S'|))$ intervals, so the overall query time is $\tilde{\mathcal{O}}(|S'|^\epsilon + |S'|^{1-\epsilon})$ which is dominated by $\tilde{\mathcal{O}}(|S'|^{1-\epsilon})$. The time for maintaining the data structure for all the intervals, as previously discussed, is $O(n^\epsilon)$. □

## 6.2 Substring Suffix Array

In this section, we show how to use our data structure to enable lookups in the suffix array of a given substring. Namely: Given three integers $i, j, k$, return $SA_{i,j}[k]$ with $SA_{i,j}$ being the suffix array of $S[i \ldots j]$.

For this purpose, we maintain the lexicographic order of suffixes starting in exponentially-increasing sized intervals in $S$ as in the previous section.

Upon query, we partition the substring $S' = S[i \ldots j]$ to $O(\log(|S'|))$ intervals. We insert a text update $S[j+1] \leftarrow \$$, where $\$ \notin \Sigma$ and $\$ >_L \sigma$ for every $\sigma \in \Sigma$. This update is required to ensure that the order of the suffixes $S$ starting in the intervals of the partition have the same lexicographic order as their respective prefixes that are suffixes of $S[i \ldots j]$. We only apply the update on the suffix select data structures for the intervals in the partition of $S'$.

For every interval $I$ in the partition, we query the lexicographical median $m_I$ of the suffixes starting within $I$. We proceed to sort the medians and find a $\frac{1}{4}$-good pivot among them similarly to the pivot finding procedure in the dynamic suffix array algorithm in 5.

Denote the pivot as $i_p$. We can find how many suffixes in $S' = S[\ldots j]$ are lexicographically smaller than $i_p$ by binary searching for $S[i_p \ldots n]$ in the sorted suffixes data structure of every interval. The index in which the binary search terminates is the amount of suffixes that are lexicographically smaller than $S[i_p \ldots n]$ starting in the corresponding interval. Summing the indices obtained from the intervals will yield the lexicographic rank $r$ of $i_p$ among the suffixes of $S'$.

We compare $k$ to the index of $r$ and continue accordingly in a binary search manner. During the search, we maintain the *lexicographic* interval $[a_I \ldots b_I]$ that may still contain $SA_{i,j}[k]$ in the suffixes interval $I$. After finding the pivot $i_p$ and its lexicographic rank $r$ among the suffixes of $S'$, we update the *lexicographic* interval of every suffixes interval $I$ to either $[a_I \ldots r_I(i_p)]$ or $[r_I(i_p) \ldots b_I]$, depending on the whether $k < r$ or $k > r$. $r_I(i_p)$ denotes the amount of suffixes starting in $I$ that are lexicographically smaller than $i_p$. The medians for the next iteration of the binary search is the median of the updated *lexicographic* interval of $I$.

**Complexity:** The procedure consists of finding $O(\log(n))$ medians in every iteration of the binary search and finding the index of $i_p$. Using our data structure, both can be done in $\tilde{O}(|I|^{1-\epsilon})$ per interval. Which is dominated by $\tilde{O}(|S'|^{1-\epsilon})$. Multiplied by the amount of iterations in the binary search, we are left with $\tilde{O}(|S'|^{\epsilon})$. We also execute an update operation to the participating intervals in order to append the $\$$ after $S'$ (and another update operation for undoing this after we output the query). So the overall query complexity is $\tilde{O}(|S'|^{\epsilon} + |S'|^{1-\epsilon})$. Maintaining the data structures when $S$ is modified can be done in $O(n^{\epsilon})$ as discussed in Section 6.1. To conclude, we get the following:

**Theorem 2.** *For every $0 \leq \epsilon \leq \frac{1}{2}$, A dynamic text $S[1 \ldots n]$ can be maintained in $\tilde{O}(n^{\epsilon})$ time per substitution to support Internal Suffix Array queries in time $\tilde{O}(|S'|^{1-\epsilon})$ with $S'$ being the queried substring.*

### 6.3 Dynamic Burrows-Wheeler Transform

The Burrows-Wheeler transform is a well known permutation of the symbols of a text. Roughly speaking, if a text has a lot of repetitions, the BWT of the text has a short run length encoding. making the BWT useful for compression.

The $i$'th symbol of the Burrows-Wheeler transform of a given text can be directly evaluated using a single lookup in the suffix array. Therefore, the following directly follows from our main result:

**Theorem 3.** *For every $0 \leq \epsilon \leq 1$, A dynamic text $S[1 \ldots n]$ can be maintained in $\tilde{\mathcal{O}}(n^\epsilon)$ time per substitution to report $BWT[i]$ queries in $\tilde{\mathcal{O}}(|n|^{1-\epsilon})$ time.*

Similar complexity can be achieved for computing $BWT[i]$ of a given substring in a dynamic text via the result of Subsection 6.2.

**Theorem 4.** *For every $0 \leq \epsilon \leq \frac{1}{2}$, A dynamic text $S[1 \ldots n]$ can be maintained in $\tilde{\mathcal{O}}(n^\epsilon)$ time per substitution to report $BWT_{S'}[i]$ in time $\tilde{\mathcal{O}}(|S'|^{1-\epsilon})$ with $BWT_{S'}$ being the BWT of a substring $S' = S[ij]$.*

### 6.4 Dynamic LCP Array

The LCP Array is a data structure that is often used alongside the suffix array in string algorithms. The $i$'th entry of the LCP Array $H[1 \ldots n]$ of a text $S$ is the LCP of the suffixes starting in $SA_S[i]$ and $SA[i-1]_S$ for $i \geq 2$ ($H[1]$ is undefined).

**Theorem 5.** *The LCP Array $H$ of a dynamic string can be maintained in $\tilde{\mathcal{O}}(n^\epsilon)$ time per update with $O(n^{1-\epsilon})$ lookup time for $H[i]$.*

**Proof:** Directly from our main result, by setting $k = n^\epsilon$ and using a dynamic LCP query data structure. Computing $H[i] = LCP([SA[i], SA[i+1])$ requires two suffix array lookups and one LCP query. □

### 6.5 Dynamic Suffix Tree

The Suffix Tree is one of the most frequently used data structures in string algorithms. Given a string $S$, the Suffix Tree of $S$ denoted as $ST_S$ is a compact trie containing all the suffixes of $S$. Every node $V$ in $S$ is associated with a string $L(V)$ that is the concatenation of the substrings written on the edges in the route from the root of $ST_S$ to $V$. It is a known fact that the leaves of the substree of $ST_S$ rooted in a node $V$ form a consecutive interval of suffixes in $SA_S$. Therefore, $V$ can be represented by two indices $i, j$ in the suffix array, such that $SA[i], SA[i+1] \ldots, SA[j-1], SA[j]$ are the leaves of the subtree rooted in $V$. We call this representation the *Suffix Array representation of $V$*. Using this representation, we can obtain the following:

**Theorem 6.** *Given a dynamic string $S[1 \ldots n]$, a data structure can be maintained in $\tilde{\mathcal{O}}(n^\epsilon)$ time per update to support the following queries in $\tilde{\mathcal{O}}(n^{1-\epsilon})$ time:*

1. **Input:** *two indices $i, j$ representing a substring $A = S[i \ldots j]$.*
   **Output:** *a Suffix Array representation of the node $V$ with $L(V) = A$*
2. **Input:** *a Suffix Array representation $(i, j)$ of a node $V$ in $ST_S$, and a symbol $\sigma \in \Sigma$.*
   **Output:** *a Suffix Array representation of the child of $V$ following the edge emerging from $V$ with a label starting with $\sigma$ (if such an edge exists).*
3. **Input:** *a Suffix Array representation $(i, j)$ of a node $V$ in $ST_S$.*
   **Output:** *a Suffix Array representation of the parent of $V$ in $ST_S$.*

   **Proof:**

1. The node $V$ in $ST_S$ for which $L(V) = S[i \ldots j]$ is the ancestor of all the suffixes starting with $S[i \ldots j]$. We can run a binary search on the suffix array for the indices of the lexicographically smallest and largest suffixes having a prefix that is equal to $S[i \ldots j]$, respectively denoted as $i'$ and $j'$. We output $(i', j')$.
2. We query the suffix array for $SA[i] = a$ and $SA[j] = b$. We use an LCP query to get $LCP(S[a \ldots n], S[b \ldots n]) = l$. It can be easily verified that $L(V) = S[a \ldots a+l-1]$. We proceed to binary search the suffix array for the indices of the lexicographically minimal and maximal suffixes starting with $S[a \ldots a + l - 1]\sigma$ respectively denoted as $i'$ and $j'$. If there are no suffixes starting with that string, report non existing edge. Otherwise, return $(i', j')$.
3. If $(i, j) = (1, n)$, then the input to the query is the root of $ST_S$ and it has no parent. Otherwise, assume that $i \neq 1$ **and** $j \neq n$. We query the suffix array for $SA[i] = x$ $SA[i - 1] = a$ and $SA[j + 1] = b$. Note that $S[x \ldots n]$ is a descendant of $V$ in $ST_S$, $S[a \ldots n]$ is the rightmost leaf in $ST_S$ that is to the left of $V$, but is not a descendant of $V$, and $S[b \ldots n]$ is the leftmost leaf in $ST_S$ that is to the right of $V$ but is not a descendant of $V$. Therefore, the parent of $V$ is either the lowest common ancestor of the leaves corresponding to $S[x \ldots n]$ and $S[a \ldots n]$ or the lowest common ancestor of the leaves corresponding to $S[x \ldots n]$ and $S[b \ldots n]$. It is a well known fact the lowest common ancestor $U$ of two leaves corresponding to suffixes $S[i_1 \ldots n], S[i_2 \ldots n]$ in the suffix tree has $L(U) = S[i_1 \ldots i_1 + LCP(i_1, i_2) - 1]$. We use LCP queries to find $l_a = LCP(a, x)$ and $l_b = LCP(b, x)$. It can be easily verified that the parent $P$ of $V$ has $L(P) = S[x \ldots x + max(l_a, l_b) - 1]$. We find this substring using $LCP$ queries and employ 1 to find the Suffix Array representation of $P$.

For every query we use (at most) a logarithmic number of suffix array lookups and LCP queries. The desired complexities are achieved by setting $k = n^\epsilon$ to the data structure of our main result. □

Theorem 6 allows performing a traversal on the Suffix Tree of a dynamically changing text, Essentially providing a dynamic variant of the suffix tree.

## 7 Conclusions and Open Problems

We presented the first dynamic algorithm for maintaining the suffix array $(SA)$ and inverse suffix array $(iSA)$ of a dynamic text. We use a new data structure which we call the $k$-words tree. For every $0 \leq \varepsilon \leq 1$, our data structure reports $SA[i]$ in time $\tilde{\mathcal{O}}(n^\varepsilon)$ and handles text modification in time $\tilde{\mathcal{O}}(n^{1-\varepsilon})$.

Our data structures enables solving several types of string queries of a dynamically changing string that could not be solved hitherto. Examples are finding the Lyndon root of a query substring, finding the suffix array of a query substring, and evaluating the $i$th symbol of the Burrows-Wheeler transform of a string or a query substring, all for a dynamically changing string.

While our algorithm gives a tradeoff between the text modification and lookup time, if we set them to be equal we get an $\tilde{\mathcal{O}}(\sqrt{n})$ time for both modification and query. We did not prove a lower bound but we believe that better bounds can be achieved.